\documentclass[12pt,authoryear]{elsarticle}
\usepackage{multirow}
\usepackage{url}
\usepackage{graphicx}
\usepackage{amssymb}
\usepackage{natbib}

\usepackage{lineno}
\usepackage[top=2cm, bottom=2cm, left=1.5cm, right=1.5cm]{geometry}

\usepackage[symbol]{footmisc}

\begin{document}

\begin{frontmatter}

\title{Practical X-ray Gastric Cancer Diagnostic Support Using Refined Stochastic Data Augmentation and Hard Boundary Box Training}

\author[inst1]{Hideaki Okamoto}
\author[inst1,inst2]{Quan Huu Cap}
\author[inst3]{Takakiyo Nomura}
\author[inst3]{Kazuhito Nabeshima}
\author[inst3]{Jun Hashimoto}
\author[inst1]{Hitoshi Iyatomi$^\ast$}{\corref{mycorrespondingauthor}}
\cortext[mycorrespondingauthor]{Corresponding author}
\ead{iyatomi@hosei.ac.jp}

\affiliation[inst1]{organization={Department of Applied Informatics, Graduate School of Science and Engineering, Hosei University},
            addressline={3-7-2 Kajino}, 
            city={Koganei},
            postcode={184-8584}, 
            state={Tokyo},
            country={Japan}}

\affiliation[inst2]{organization={AI Development Department, Aillis Inc.},
            addressline={2-2-1 Yaesu}, 
            city={Chuo},
            postcode={104-0028}, 
            state={Tokyo},
            country={Japan}}

\affiliation[inst3]{organization={Department of Radiology, Tokai University School of Medicine},
            addressline={143 Shimokasuya}, 
            city={Isehara},
            postcode={259-1193}, 
            state={Kanagawa},
            country={Japan}}

\begin{abstract}
Endoscopy is widely used to diagnose gastric cancer and has a high diagnostic performance, but it must be performed by a physician, which limits the number of people who can be diagnosed.
In contrast, gastric X-rays can be taken by radiographers, thus allowing a much larger number of patients to undergo imaging.
However, the diagnosis of X-ray images relies heavily on the expertise and experience of physicians, and few machine learning methods have been developed to assist in this process.
We propose a novel and practical gastric cancer diagnostic support system for gastric X-ray images that will enable more people to be screened.
The system is based on a general deep learning-based object detection model and incorporates two novel techniques: refined probabilistic stomach image augmentation (R-sGAIA) and hard boundary box training (HBBT).
R-sGAIA enhances the probabilistic gastric fold region and provides more learning patterns for cancer detection models.
HBBT is an efficient training method that improves model performance by allowing the use of unannotated negative (i.e., healthy control) samples, which are typically unusable in conventional detection models.
The proposed system achieved a sensitivity (SE) for gastric cancer of 90.2\%, higher than that of an expert (85.5\%).
Under these conditions, two out of five candidate boxes identified by the system were cancerous (precision = 42.5\%), with an image processing speed of 0.51 seconds per image. 
The system also outperformed methods using the same object detection model and state-of-the-art data augmentation by showing a 5.9-point improvement in the F1 score.
In summary, this system efficiently identifies areas for radiologists to examine within a practical time frame, thus significantly reducing their workload.
\flushleft {\it Declarations of interest}: None
\end{abstract}

\begin{graphicalabstract}
\includegraphics[width=150mm]{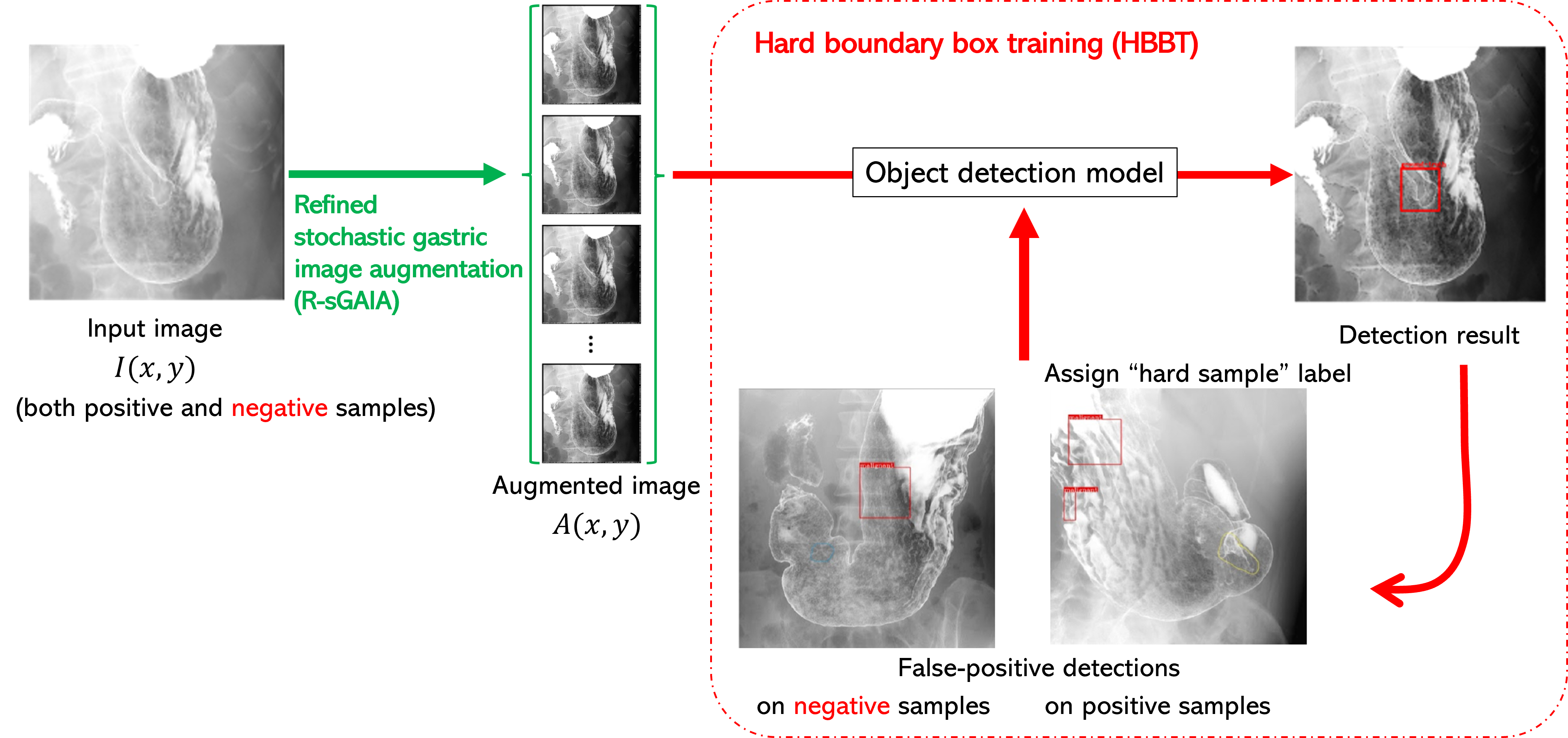}
\end{graphicalabstract}

\begin{highlights}
\item A novel gastric cancer detection system using X-rays enables scalable screening.
\item R-sGAIA enhances gastric folds, providing diverse patterns for cancer detection. 
\item HBBT enables training with unannotated data, reducing false positives effectively.
\item The system outperforms experts with 90.2\% sensitivity, diagnosing at 0.51s per image.

\item Our R-sGAIA and HBBT are available at \url{https://github.com/IyatomiLab/RsGAIA_HBBT/}.
\end{highlights}


\end{frontmatter}


\section{Introduction}
\label{sec:intro}
Gastric cancer is the third most deadly cancer globally and has a poor prognosis, especially in advanced stages \citep{Rawla2019}.
However, early detection before metastasis can significantly improve outcomes, which makes early diagnosis and appropriate treatment crucial.
Gastric cancer is primarily diagnosed through endoscopy or radiography.
Endoscopy results in a sensitivity (SE) of 95.4\%, which is superior to other methods, such as gastric radiography or X-ray, at 85.5\% \citep{Hamashima2013}.
However, endoscopy can only be performed by a physician, and the time and cost involved in the procedure limit the number of patients who can be screened.
%
In contrast, gastric X-rays can be taken by radiographers, allowing them to examine many patients in a shorter time.
This makes the method suitable for mass screening and provides a significant advantage over endoscopic diagnosis.
In Japan, a mass screening system was established, and, in 2019, at least 3.87 million people underwent gastric X-rays for gastric cancer screening.
Of these, 183,000 people were indicated for further testing, and 2,553 cases of gastric cancer were ultimately detected. By comparison, only 374,000 people underwent endoscopy for the same purpose \citep{Statistics2019}.
Although endoscopy’s effectiveness has led to a shift toward this diagnostic method, developing superior diagnostic support technology for gastric radiography would significantly contribute to the early detection and reduction of mortality from gastric cancer.

Gastric X-ray images are later interpreted by doctors to diagnose the stomach’s shape and mucosal atrophy \citep{Uemura2001, epidemiology2002, Correa2004}.
An additional advantage of X-rays over endoscopy is their ability to reveal minute irregularities indicative of stage IIc lesions (early-stage tumors with slight indentations), which appear as a subtle elevation within the indentation that lies beneath it.
However, accurately reading these images requires a high level of skill, which many gastrointestinal radiologists may lack. When performed by a skilled physician, there is a high likelihood of assessing the extent of cancerous invasion deep within the surface of an early-stage cancer based on changes in the microscopic structure of the surface.
The limitations of this diagnostic method lie in its dependence on extensive physician experience and the particular difficulty of detecting early-stage cancers, which leads to a lower cancer detection capacity than that of endoscopy \citep{Hibino2023}.
Therefore, there is a pressing need for an automated diagnosis assistance system that can accurately and promptly detect gastric abnormalities from gastric X-ray images in clinical settings.

Healthy gastric folds are thin, smooth, and parallel; however, their thickness and size can be altered by gastric conditions such as inflammation, infection, ulcers, and tumors that lead to irregular bends, depressions, and tears in the lining mucosa \citep{Correa2004}.
Most research on diagnostic support using gastric X-rays has focused on detecting gastritis, the precursor to gastric cancer, or the presence of {\it Helicobacter pylori} infection, which is believed to be a major cause of gastric cancer \citep{Cooper1990, Kita1992, Minemoto2010, Nagano2010, Abe2014, Ishihara2015, Ishihara2017, Togo2019, Kanai2019, Kanai2020}.
Among these studies, methods based on deep learning, particularly convolutional neural networks (CNNs), have shown excellent results in detecting gastritis.
The main advantage of these deep learning techniques is their ability to automatically capture efficient features for the target task, which is promising in areas where disease features are difficult to define.
However, studies directly detecting gastric cancer from radiographs are limited, despite the significance of gastric cancer diagnosis in terms of life expectancy. This may be because gastric cancer, especially in its early stages, is more difficult to diagnose compared to gastritis, where lesions often expand and develop to some degree. Reportedly, even with the latest machine learning techniques, distinguishing between gastritis and gastric cancer on endoscopic images—where signs of the disease are more visible--remains challenging \citep{Horiuchi2020}.

In the field of computer vision, CNN-based object detection methods (e.g., Faster R-CNN \citep{fasterrcnn}, YOLOv3 \citep{yolov3}, EfficientDet \citep{efficientdet}, and YOLOv10 \citep{yolov10}), which can simultaneously detect and classify the location of multiple objects in images, have been proposed and widely used.
These methods are particularly valuable in medical applications because they not only detect regions of interest but also provide high explanatory power for the predicted results, as opposed to simply classifying images as benign or malignant \citep{Hirasawa2018, Li2018, Okamoto2019, Laddha2019, Zhang2019}.
For instance, in a study on gastric X-ray images, \citet{Laddha2019} used YOLOv3 to automatically detect gastric polyps and achieved a mean average precision of 0.916.
However, these object detection methods require precise annotation of lesion locations prior to training, which is extremely costly. To avoid high annotation costs while still benefiting from the explanatory advantages of object detection models, previous studies on gastritis detection \citep{Ishihara2017, Togo2019, Kanai2019, Kanai2020} used patch-based CNNs, which only required image-level annotations.
In the context of gastric cancer, where diagnostic challenges and outcomes are more severe, it is critical that physicians are able to interpret the results provided by diagnostic support systems.

With this motivation, we developed a diagnostic support system that directly targets gastric cancer using X-ray images,which have received little attention in the literature, and offers high explanatory power for the results \citep{Okamoto2019}.
Thanks to their gastric image augmentation (sGAIA) developed based on medical knowledge, this system accurately represented candidate regions of gastric cancer within bounding boxes. In an evaluation using images from patients not included in the training set, the recall (SE) and precision of the detected bounding boxes were 92.3\% and 32.4\%, respectively.
This recall is about 7\% higher than that achieved by doctors (85.5\%), and a box-by-box evaluation confirmed a true-to-false detection ratio of 1:3, which is sufficient for practical use.

While the object detection model trained with sGAIA reached a practical level of accuracy, further performance improvements are desirable.
Currently, widely used object detection methods have the disadvantage of not being able to explicitly use negative (i.e., healthy control) data for training, in addition to the high cost of annotation. When such models are used for lesion detection, only images with lesions are used for training, while control images without lesions are excluded.
This limitation presents an opportunity for improvement, which is a key focus of this study.
In medical imaging, data are expensive, and we believe that actively using healthy control images for training can enhance the model's overall ability to detect lesions, thereby effectively reducing annotation costs while maintaining performance.

Therefore, in this study, we propose a practical gastric cancer diagnostic support system that includes two new techniques: refined stochastic gastric image augmentation (R-sGAIA), an improvement of sGAIA \citep{Okamoto2019}, and hard boundary box training (HBBT), which enables the use of negative or healthy control images for training, a feature not previously utilized in object detection neural networks.
To suppress high-confidence false positives predicted by the object detector, HBBT registers each of these boxes as a ‘‘hard sample’’ class and retrains the model until predetermined convergence conditions are met.
We investigated 4,724 gastric X-ray images obtained from 145 patients in a clinical setting and confirmed the practicality and effectiveness of the proposed system by evaluating gastric cancer detection performance using a subject-based five-fold cross-validation strategy.

The contributions of this study are as follows:
\begin{itemize}
\item We propose a novel gastric cancer diagnostic support system for gastric X-rays that can be captured by radiographers.
The use of radiographers rather than physicians allows more people to be screened than the current mainstream use of endoscopy, which can only be performed by physicians.
The system is based on a general deep learning-based object detection model and includes two novel technical proposals: R-sGAIA and HBBT.
\item R-sGAIA is a probabilistic gastric fold region enhancement method used to provide more learning patterns for cancer detection models.
\item HBBT is an efficient and versatile training method for common object detection models that allows the use of unannotated negative (i.e., healthy control) samples that otherwise cannot be used for training in conventional detection models.
This approach reduces false positives and thus improves overall performance.
\item The SE of the proposed system for gastric cancer (90.2\%) exceeds that of the expert (85.5\%), while 42.5\% of the detected candidate box regions show cancerous lesions, with a processing time of 0.51 seconds/image. 
It is 5.9 points higher on the F1 score compared to methods that that use the same object detection model and state-of-the-art data augmentation.
In short, the system efficiently shows the radiologist where to look, thereby greatly reducing the radiologist's workload.
\item
Our R-sGAIA and HBBT are abailable at \url{https://github.com/IyatomiLab/RsGAIA_HBBT/}.
\end{itemize}

\section{Related work}
\subsection{Diagnostic support using gastric X-ray images}
In the 1990s, methods were proposed for effectively extracting the pattern of gastric folds by applying binarization as a preprocessing step and estimating the gastric region based on the position of the barium reservoir \citep{Cooper1990, Kita1992}.
In the 2010s, more empirical studies began to focus on gastric folds \citep{Minemoto2010, Nagano2010, Abe2014, Ishihara2015}. 
\citet{Ishihara2015} calculated numerous statistical features (7,760 per case) for {\it H. pylori} infection, a major cause of gastric cancer, by analyzing post-infection mucosal patterns and gastric folds.
Their support vector machine, trained on data from 2,100 patients (with eight images per person), achieved an SE of 89.5\% and a specificity (SP) of 89.6\%.

More recently, deep learning techniques have been applied to the detection of gastritis using patch-based CNNs on gastric X-ray images \citep{Ishihara2017, Togo2019, Kanai2019, Kanai2020}.
Although these methods targeted gastritis, which is easier to diagnose than gastric cancer, they achieved high detection accuracy and were considered practical for diagnostic assistance.
\citet{Kanai2020} used a patch-based CNN to detect {\it H. pylori}-associated chronic atrophic gastritis.
They trained image patches from both the inside and outside of the stomach and made a diagnosis by taking a majority vote of each estimated result in the stomach region. Manual annotation of regions of interest (ROIs) prior to training, along with self-learning to increase the number of training patches, ultimately yielded a harmonic mean SE and SP of 95.5\% for the test data from different patients.
While excellent screening results have been achieved for gastritis, the detection of more serious and difficult-to-diagnose gastric cancers has not been well studied. In the case of gastric cancer, it is also crucial to provide an explanation for the results.
The proposed method significantly improves upon our previous work on a gastric cancer detection system that can present explainable results \citep{Okamoto2019} by incorporating two new technical advancements.

\subsection{Diagnostic support research for endoscopic images}
In recent years, many methods for analyzing narrow-band endoscopy images to assist in diagnosing gastric cancer using deep learning technology have been proposed \citep{Kanayama2019, Wu2019, Horiuchi2020, Li2020, Gong2023}.

\citet{Wu2019} used classic VGG16 and ResNet-50 models to achieve an SE of 94.0\% and an SP of 91.0\% based on 24,549 images for diagnosing early gastric cancer.
\citet{Li2020}  employed an Inception-v3 model to achieve an SE of 91.2\% and an SP of 90.6\% based on 2,429 images of gastric magnification endoscopy with narrow-band imaging.
\citet{Horiuchi2020} trained a GoogLeNet model with 1,499 images of gastric cancer and 1,078 images of gastritis, thereby achieving an SE of 95.4\%, an SP of 71.0\% as well as an area under the receiver operating characteristic (ROC) curve (AUC) of 0.852.
\citet{Gong2023} developed a convolution and relative self-attention parallel network using 3,576 endoscopic images from 205 patients and reported an F1 score of 0.948 in their experiments with a 7:3 split of training and test images.
While these methods produced excellent results, none of them explicitly stated whether the training dataset was completely separate from the patients whose images were in the evaluation set.
This raises the possibility that images from the same patients were included in both the training and test data, potentially leading to inflated performance scores due to the close similarity between the two datasets.

In contrast, \citet{Kanayama2019} constructed a diagnostic support system using 129,692 gastrointestinal endoscopy images with sophisticated CNN-based networks, including one synthesizer network and two different types of discriminator networks.
Importantly, they ensured that images from the same patient were not divided between the training and evaluation sets, thus evaluating performance appropriately.
They reported an average precision (AP) of 0.596 for cancer detection.
Although the numerical performance of this method may seem lower, it is based on a newer and more sophisticated machine learning model than those used in previously reported methods and represents a more accurate diagnostic capability.
They further enhanced their model by generating 20,000 high-resolution lesion images from 1,315 narrow-field-of-view images using their own generative adversarial network \citep{gan} and, by adding these to the training set, improved the AP in cancer detection to 0.632.
\citet{Shibata2020} constructed and rigorously evaluated a diagnostic model using Mask R-CNN on 1,208 images from 42 healthy subjects and 533 images from 93 unhealthy subjects (including gastric cancer). Their model achieved an SE of 96\% for all tumor types, a false positive rate of 0.1/image, and an F1 score of 0.71 for gastric cancer.
\citet{Ahmad2023} trained and evaluated a larger dataset than in previous studies (27,200 training and 6,800 testing images) with a novel approach that added squeeze and excitation attention blocks to YOLOv7 \citep{yolov7}, thereby achieving an F1 score of 0.71 (precision = 72\%, recall = 69\%). 
While it is unclear whether proper data partitioning was employed for this method, it is important to note that the number of images used was significantly larger than in other studies. 
Additionally, the use of more sophisticated methods yielded results different from earlier studies (which reported F1 scores around 90\% or higher) where substantial data leakage was suspected. 
Given that the results were comparable to other reports \citep{Shibata2020} with clearly distinct data partitions, it can be inferred that the data division was likely appropriate.

\subsection{Hard negative mining}
Hard sample or hard example mining, a technique that improves model performance by explicitly retraining the model on error-prone or hard-to-identify data (hard samples), is widely used in machine learning \citep{Felzenszwalb2009, Shrivastava2016}.
It is particularly effective in addressing class imbalance and has been extensively applied, alongside metric learning, in tasks such as face or person detection from complex backgrounds \citep{Zhang2016, Smirnov2018, Chen2020}, among other applications.

In recent years, deep learning-based object detection models have also adopted this strategy and reported significant performance improvements by relearning misdetected bounding box regions, including in the medical field \citep{Li2019, Tang2019}.
However, such object detection models typically cannot utilize images that do not contain any detection targets (such as background images or healthy control images) for training because the error function to be minimized is defined based on the position and size of the bounding box for each detection target.
Existing hard sample mining methods have been applied to data with at least one annotation box of the target class and have used overdetected or undetected regions or image patches as hard samples.
Our HBBT is a form of hard sample mining or hard negative mining that retrains regions of false positives. 
However, it differs significantly in that it allows the use of healthy control cases for training, which could not be used in traditional object detection models.
This is a particularly important contribution in the medical field, where the cost of acquiring positive samples is high.

\section{Practical gastric cancer diagnostic support system}
The system highlights suspected cancerous sites with bounding boxes, thus allowing physicians to confirm the validity of the estimations.
The two technical innovations in this study significantly improved system performance and achieved practical accuracy. The overall structure of the proposed system is shown in Figure 1. This study was approved by the Institutional Review Board of Tokai University Hospital (No. 20R-033, approval date: June 12, 2020).
The main technical contributions were (1) R-sGAIA, which generates a diverse set of images that effectively emphasize the gastric fold region for data augmentation, and (2) HBBT, which reduces false positives by assigning ‘‘hard sample’’ classes to bounding box regions where false positives are detected and iteratively learning from them.
The object detection model used is EfficientDet \citep{efficientdet}, which incorporates EfficientNet \citep{efficientnet} as its backbone, a CNN known for its excellent performance, although the choice of model was arbitrary.
The input images were 2,048 $\times$ 2,048 pixels, and the output consisted of a bounding box surrounding the presumed cancerous region in the image along with its confidence score.

\begin{figure*}[t]
\begin{center}
\includegraphics[width=150mm]{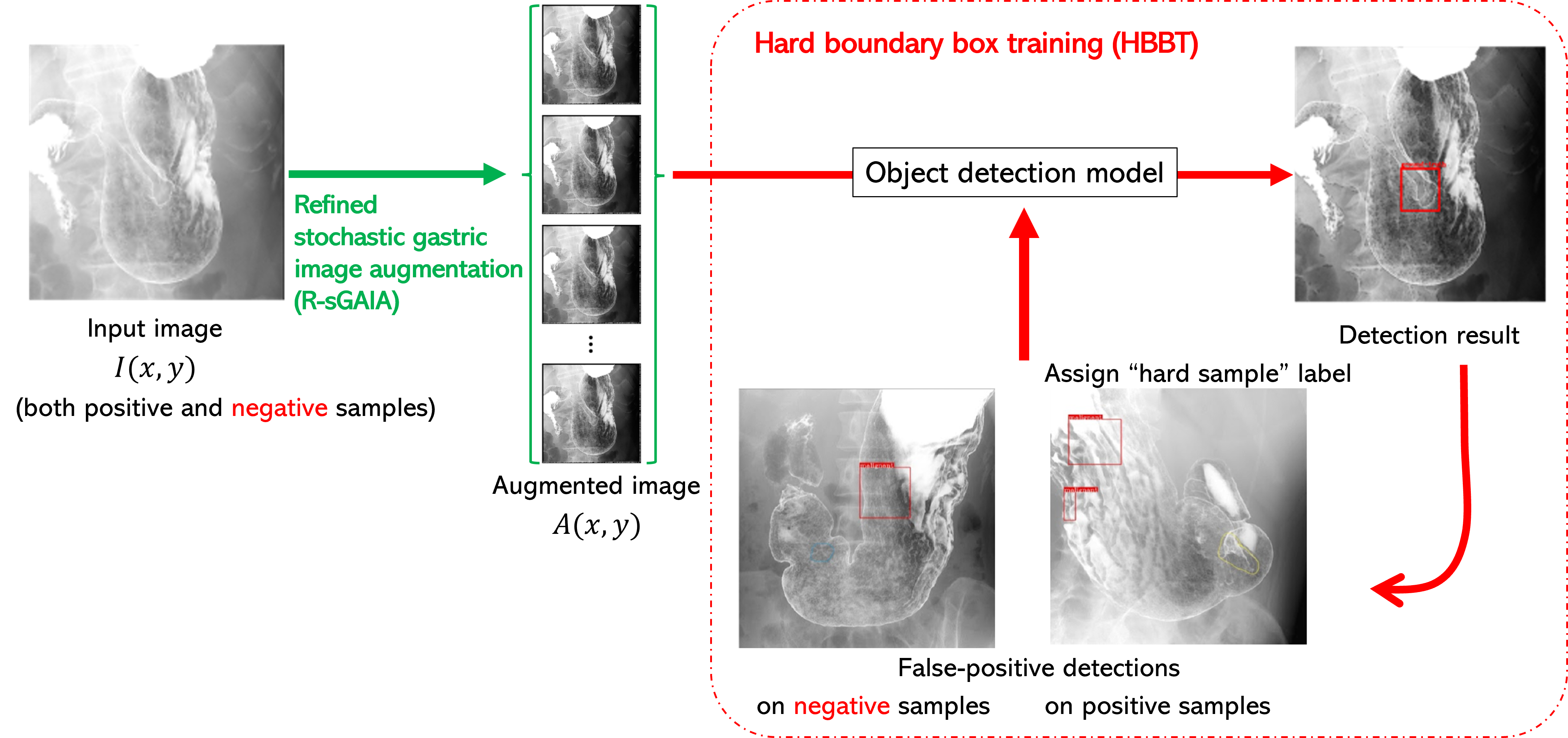}
\caption{Overview of the proposed gastric cancer diagnostic support system.}
\end{center}
\end{figure*}

\begin{figure*}[t]
\begin{center}
\includegraphics[width=150mm]{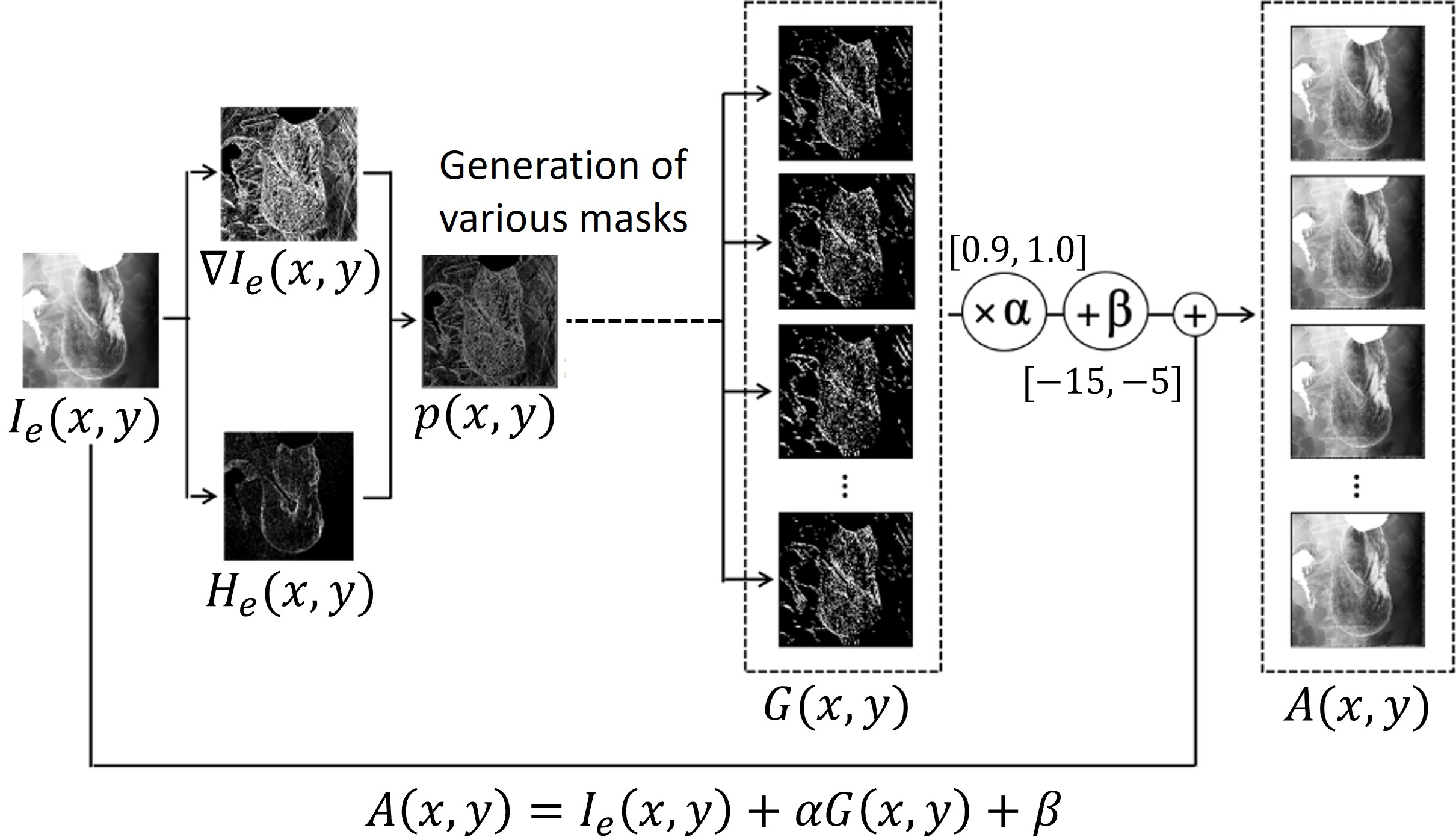}
\caption{Overview of refined stochastic gastric image augmentation (R-sGAIA).}
\end{center}
\end{figure*}

\subsection{R-sGAIA}
The proposed R-sGAIA is a refined version of our previous method, sGAIA \citep{Okamoto2019}.
Both R-sGAIA and sGAIA are online data augmentation techniques developed based on medical knowledge to generate diverse images that emphasize gastric folds in X-ray images, aiming to detect inflammation on the gastric mucosal surface.
The original sGAIA improved gastric cancer detection by 6.9\% in terms of the F1 score (recall 92.3\%, precision 32.4\%) when applied to a Faster R-CNN model with ResNet-101 as the backbone, thereby validating its effectiveness for practical decision support.
However, in sGAIA, the emphasis probability of each pixel $p(x,y)$, crucial for effective enhancement, was subjectively determined as a discrete value based on preliminary experiments.
In the proposed R-sGAIA, $p(x,y)$ is instead modeled as a flexible continuous function determined by only two hyperparameters.
A schematic of the proposed R-sGAIA process is shown in Figure 2.
Similar to sGAIA, R-sGAIA consists of four steps, with step 2 being the only difference.

\begin{itemize}
\item Step 1: Calculation of edge strength $E(x,y)$

First, a contrast-enhanced image $I_e(x, y)$ is generated from the original gastric X-ray image $I(x, y)$ using histogram equalization.
Then, its gradient and high-frequency components, $\nabla I_e(x, y)$, and $H_e(x, y)$, are obtained using general image processing techniques, such as Canny edge detection filter and Butterworth high-pass filter, respectively.
Each component is normalized to [0, 1], and the normalized edge strength $E(x, y)$ is calculated as follows:

\begin{equation}
\mbox{$E(x, y)$} = ( \mbox{$\overline{\nabla I_e(x, y)}$} + \mbox{$\overline{H_e(x, y)}$} ) / 2,
\end{equation}
where $\overline{v(x,y)}$ is the linearly normalized value of $v(x,y)$ to [0,1].

\item Step 2: Calculation of the probability of a gastric fold region $p(x, y)$

The area of the edge of the gastric folds that is subject to augmentation is determined for each pixel as its probability $p(x, y)$.
The proposed R-sGAIA differs from sGAIA in this step. 
In R-sGAIA, the edge intensity $e=E(x, y)$ of each pixel is used to obtain a probability map $p(x,y)$ of the gastric folds and edges using probabilities based on the sigmoid function defined below.
\begin{eqnarray}
\mbox{$p(x,y)$} &=& \mbox{$p(e)$} = \mbox{$p(E(x,y))$} \nonumber \\
                &=& \frac{1}{ 1 + \exp (-\gamma( E(x,y) - \theta))}.
\end{eqnarray}
The parameters $\gamma$ and $\theta$ are hyperparameters that define the slope of the sigmoid function and adjust the probability of running the augmentation of $E(x,y)$ to be 50\%.
They are determined based on the diagnostic performance of the validation data, as described below.
Figure 3 compares the enhancement selection probability $p(e=E(x,y))$ based on the feature intensity $e$ for R-sGAIA and sGAIA.
Note that, in sGAIA, the probability $p(e)$ is subjectively determined based on preliminary experiments, as described above.

\item Step 3: Determination of gastric fold edge region $G(x,y)$

According to the probability map $p(x,y)$, a binary mask $G(x,y)$ of the gastric fold regions to be enhanced is determined by pixel-by-pixel sampling.
Since this result is noisy, a morphological open operation is performed to remove many small isolated regions.
$G(x, y)$ is created in such a way that a different mask is generated for each trial. 
This is the key to generating enhanced images of different gastric folds.

\item Step 4: Generation of enhanced gastric fold images $A(x, y)$

Finally, an enhanced gastric fold edge image $A(x, y)$ is obtained as follows:

\begin{equation}
\mbox{$A(x, y)$} = \mbox{$I_e(x, y)$} + \alpha \, \mbox{$G(x, y)$} + \beta.
\end{equation}
\end{itemize}

Here, $\alpha$ and $\beta$ are augmentation parameters.
The range of $\alpha$ was from 0.9 to 1.0, and $\beta$ was from $-$15 to $-$5, as determined by prior experiments.
During runtime, these parameters are randomly selected from within their respective ranges.
Using these parameters, the proposed R-sGAIA probabilistically enhances the intensity of the gastric fold region each time to generate a different enhanced gastric image for each trial.
This makes it highly compatible with online data augmentation in deep neural network training. Another significant advantage of R-sGAIA is that it does not require any additional runtime.
We applied this R-sGAIA for online augmentation to train a machine learning model (EfficientDet-D7) used for cancer region detection.

\begin{figure}[t]
\begin{center}
\includegraphics[width=150mm]{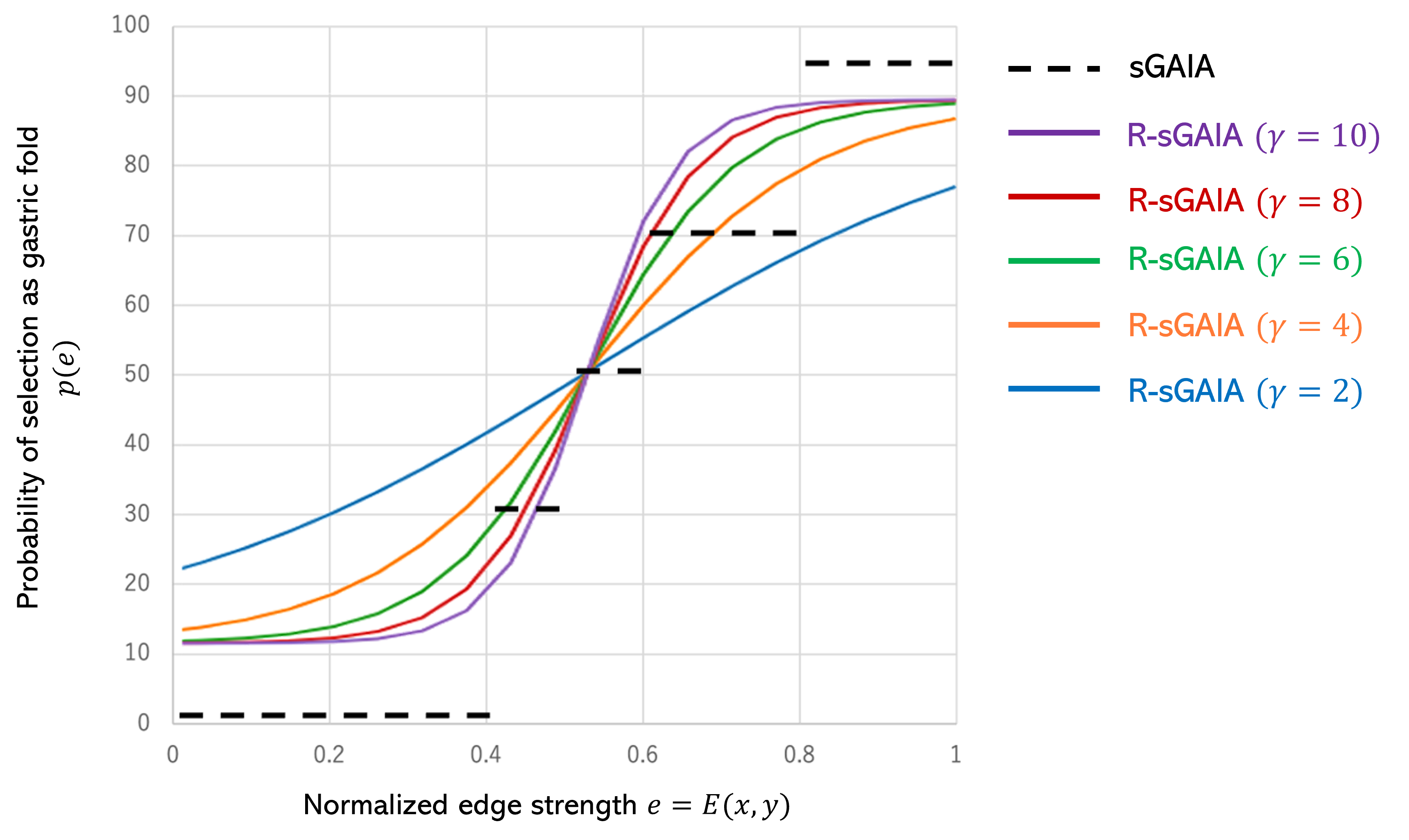}
\caption{Probability of being detected as a gastric fold edge to be highlighted.}
\end{center}
\end{figure}

\begin{figure}[t]
\begin{center}
\includegraphics[width=150mm]{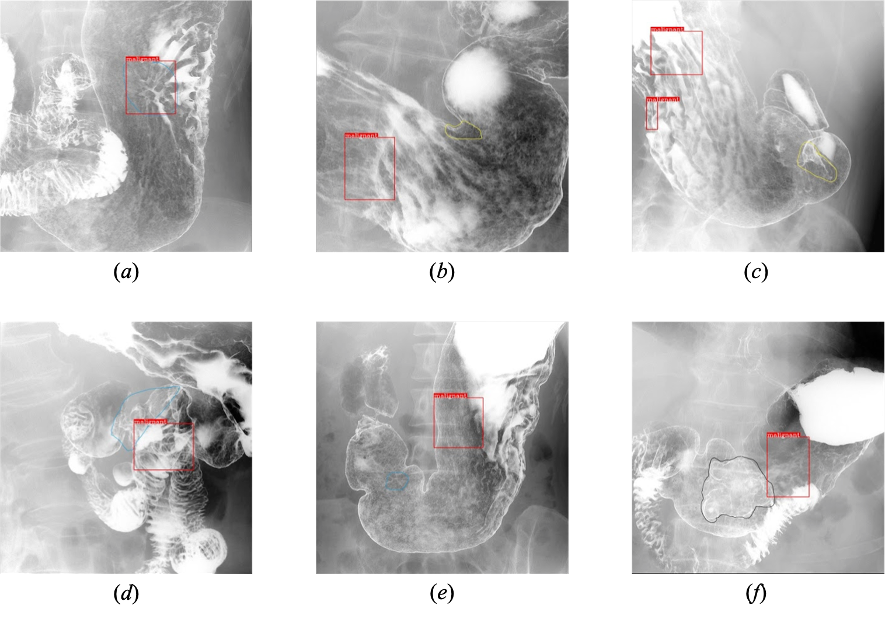}
\caption{Positive example and five typical false positives provided by experts (as shown in red box with ``malignant" label. In HBBT, these false detection regions are labeled as the ``hard boundary box" class, and the model is retrained accordingly.). \newline (a) Positive example.  (b) The area overlapping the osteophytes and spinous processes of the vertebral body.  (c) Area where folds are gathered due to lack of air in the stomach. (d) Area where the duodenal Kerckring’s folds overlap with the stomach. (e) Areas where the mucosal surface of the stomach is irregular due to chronic gastritis. (f) Areas of advanced cancer that appear as wall changes rather than masses.}
\end{center}
\end{figure}

\subsection{HBBT}
The task of detecting cancerous regions in gastric X-ray images is a one-class detection problem that focuses exclusively on positive (i.e., malignant) regions.
However, due to the presence of numerous regions with similar visual features, distinguishing these regions is challenging not only for machine learning models but also for experts.
Figure 4 illustrates typical false detections identified in experiments using the EfficientDet-D7 object detection model.
The results indicate that most of these false detections can be categorized into five categories, as shown in examples (b)--(f).
However, it is difficult to predefine and control these features explicitly.
Reducing such hard negatives--non-target regions with high visual similarity--is crucial.
As previously mentioned, existing deep learning-based object detection algorithms generally lack the capability to explicitly learn from readily available negative data, which presents a significant limitation, particularly in medical tasks such as this.

The concept of HBBT is illustrated on the right side of Figure 1.
HBBT extends the concept of hard negative mining by more effectively identifying regions prone to false negatives.
It achieves this by incorporating healthy data that could not previously be utilized for training and retraining the model to avoid detecting these regions.
In essence, HBBT leverages healthy data that was previously unavailable for training to reduce false negatives in regions where the model might otherwise misclassify with high confidence, thereby improving detection accuracy.
The concrete procedure of HBBT for any object detection model $M$ is as follows:
\begin{itemize}
\item Step 1: Train the model $M$ using the original labeled training data where the ground truth bounding boxes correspond to the target malignant label.
\item Step 2: Perform the box detection procedures using model $M$ on healthy (non-target) data. Each falsely detected bounding box region (e.g., as shown in Figure 4, (b)--(f)) is labeled as a ``hard sample" class. We refer to these false detection boxes as hard boundary boxes, and the images containing them are subsequently added to the training data.
\item Step 3: The model $M$ is re-trained with updated training data for the two-class detection task of ``malignant” and ``hard samples.”
\end{itemize}
Note that the ``hard sample" class represents a non-cancerous class and is ignored during the execution of model $M$, thus treating its output as negative.

\section{Experiments}
\subsection{Datasets and evaluation criteria}
The image set used in this study consisted of 4,724 gastric X-ray images (1,117 images with 1,504 lesion annotations by radiologists and 3,607 images without lesions) from 145 patients in clinical settings provided by the Tokai University School of Medicine, Japan.
A total of 116 images from five randomly selected patients (49 of them with 49 annotated boxes) were used as validation data to determine the hyperparameters and to decide when to terminate HBBT.
The remaining 4,608 images from 140 patients were used to train and evaluate the proposed system.
Please note that we are not permitted to disclose the image data used in this experiment.
Each image was an 8-bit grayscale with a resolution of 2,048 $\times$ 2,048 pixels.
Based on expert annotations of the gastric cancer locations, the smallest rectangle surrounding each location was defined as a bounding box to be detected.

A subject-based five-fold cross-validation was used for training and evaluation, with precision, recall, and F1 score serving as the performance indicators for this system.
The system provided candidate lesion bounding boxes with associated confidence scores exceeding a predefined detection threshold $\xi$.
In the subject-based five-fold cross-validation, subjects were randomly divided into five groups. Images from four groups were used for training, while images from the remaining group were used for evaluation.
This process was repeated five times, with a different group selected for evaluation each time.
In medical research, including this experiment, data from the same subject can be inherently similar.
Using standard cross-validation could lead to a virtual data leak and result in an overestimation of model performance due to the similarity between training and test data.
To mitigate this and ensure rigorous evaluation, we adopted the subject-based evaluation strategy.

\subsection{Details of the detection model}
We used EfficientDet-D7 \citep{efficientdet} as a gastric cancer detection model with an input resolution of 1,536 $\times$ 1,536 pixels.
One difference from the original image size was the margin for random cropping performed during training. 
The entire network was trained end-to-end, which allowed it to simultaneously estimate the detection target at any location and size within the input image and classify its content (i.e., ``cancer").
For details on the model, please refer to the original paper.
The learning rate was set to 0.001, with a batch size of 32 and 100 training epochs.
The optimization method used was momentum stochastic gradient descent (SGD) \citep{momentumsgd}, with the exponential decay rate of the first-order moment set to 0.9.
To construct the proposed gastric cancer detection system, we applied the proposed R-sGAIA as an online augmentation technique during training.
Additionally, healthy control images without cancer lesions, which are typically not used in general object detection models, were incorporated into the training process using HBBT.

\subsection{Experimental conditions}
\subsubsection{Determination of hyperparameters in R-sGAIA}
In this section, we describe the hyperparameters $\theta$ and $\gamma$ of the proposed R-sGAIA.
$\theta$ is a hyperparameter that determines the edge strength $E(x,y)$ for which the probability of being considered a gastric fold is 50\% in the sigmoid function $p(e=E(x,y))$ and was determined to be 0.55 based on the distribution of the validation results.
For $\gamma$, which characterizes the slope of $p(e)$, we selected $\gamma = 4$ from 2, 4, 6, 8, and 10 (Table 1) from the detection performance on the validation images.

\begin{table}[t]
\begin{center}
\caption{Detection performance in validation images for determining hyperparameter $\gamma$.}
\begin{tabular}{c|c|c|c} \hline
$\gamma$ & Precision  (\%) & Recall  (\%) & F1 score\\ \hline
2 & 35.7 & 89.6 & 0.511 \\
\textbf{4} & 39.6 & 91.7 & \textbf{0.553} \\
6 & 31.2 & 91.7 & 0.466 \\
8 & 32.8 & 89.6 & 0.480 \\
10 & 30.3 & 89.6 & 0.453 \\ \hline
\end{tabular}
\end{center}
\end{table}

\begin{figure*}[t]
\begin{center}
\includegraphics[width=160mm]{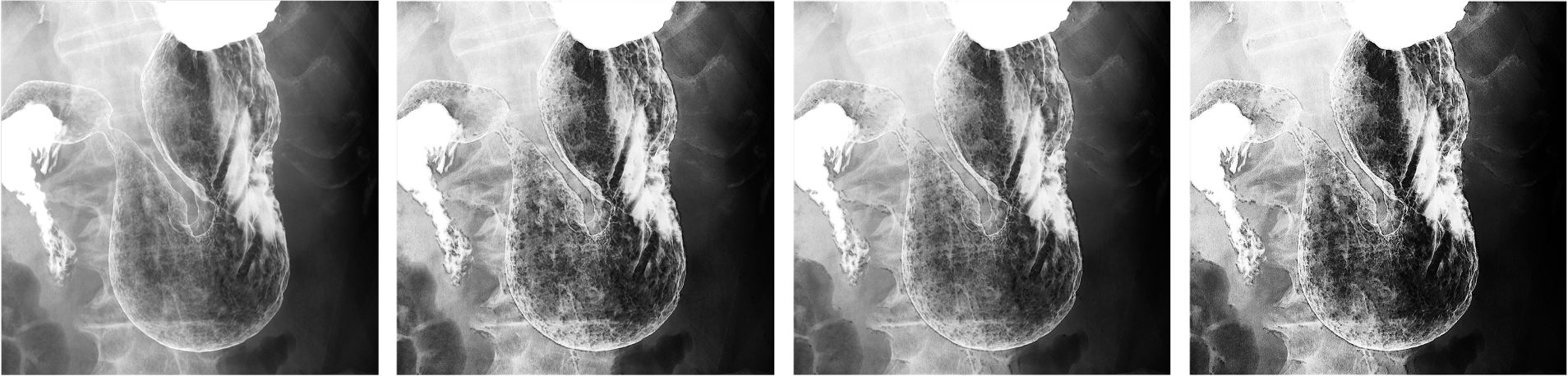}
\caption{Example of augmented images (leftmost: original, others: enhanced images with R-sGAIA).}
\end{center}
\end{figure*}

\subsubsection{Determination of hyperparameter in HBBT}
The detection threshold for hard boundary boxes in HBBT is set to match the detection threshold for positive boundary boxes across the entire model $\xi$.
In this study, the parameter $\xi$ was determined to achieve a sensitivity of approximately 90\% for detecting candidate cancer regions, as detailed below.
The primary goal of HBBT is to reduce false positives that the model detects with high confidence.
When applying HBBT, a higher detection threshold generates high-quality hard boundary boxes.
However, if the threshold is set too high, the number of hard boundary boxes becomes limited, reducing the overall effectiveness of HBBT. 
Conversely, if the threshold is set too low, it may produce an excessive number of hard boundary boxes unrelated to false positives, potentially interfering with the detection of critical cancer regions near these boxes.
In preliminary experiments where $\xi$ was set to this value, the best results were achieved when the threshold for detecting hard boundary boxes was equal to the threshold for detecting cancer regions, $\xi$.

\subsubsection{Detection of suspicious cancerous regions}
To confirm the effectiveness of the proposed system with R-sGAIA and HBBT, we conducted comparative experiments using the following combinations of conditions.
Because there is no directly comparable literature on gastric cancer detection, we implemented Laddha et al.'s (2019) method based on YOLOv3 and compared the performance on our task.

\begin{itemize}
\item Model selection

We selected two deep learning-based object detection models for this task: ResNet-101 + Faster R-CNN (model 1) and EfficientDet-D7 (model 2).
The first model has been widely used in the literature, including in sGAIA studies.
The second model is one of the advanced object detection models frequently employed in many studies, although newer models have since been proposed.
The training images used in the experiment to detect candidate cancer regions were grayscale, not color images.
Also, because there was no suitable pre-trained model available, we trained the model from scratch using He’s initial weights \citep{He_init2015} with the random seed set to ``2021".
RandAugment \citep{randaugment}, one of the state-of-the-art online augmentation methods, was applied to these models and served as the baseline for each (i.e., baseline models 1 and 2). 
Color augmentation was omitted because this task involved grayscale images.
Based on the detection performance of the validation dataset, 4 of the 13 augment types were selected with an intensity level of 3 (i.e., $M=4$ and $N=3$ as defined in the original paper).
The configuration of these machine learning models and training techniques is identical to that presented in the original paper.

%
\item Comprehensive evaluation of the proposals

We compared the performance of each model with and without the proposed techniques. For data augmentation, we used either RandAugment, conventional sGAIA (+sGAIA), or the proposed R-sGAIA (+R-sGAIA). The use of the proposed HBBT was indicated as either applied (+HBBT) or not applied.
\end{itemize}

\subsubsection{Computational environment for model training and evaluation}
In all of the experiments, we used a computer equipped with a Xeon E5-2650v4 2.2GHz CPU, 256GB of memory, and a GeForce RTX 2080 Ti GPU, running on Ubuntu 20.04 LTS.
PyTorch was used as the deep learning framework.
The implementations of Faster R-CNN\footnote{https://pytorch.org/vision/master/models/faster\_rcnn.html}, EfficientDet\footnote{https://github.com/rwightman/efficientdet-pytorch}, and RandAugment\footnote{https://pytorch.org/vision/main/generated/torchvision.transforms.RandAugment.html} were sourced from the links provided.
Please note that the OS versions listed here reflect those in use at the time of the initial submission and were not the latest versions.

\section{Results}
Figure 5 shows an example of the augmented image obtained with the proposed R-sGAIA for the leftmost original image.
Figure 6 compares cancer detection on the EfficientDet-D7 network (baseline model 2) with and without the proposed R-sGAIA and HBBT.
The input images are shown in (a), the detection results with RandAugment in (b), and those with the proposed R-sGAIA + HBBT in (c).
The proposed system (c) accurately detected the gastric cancer region without being affected by mucosal irregularities or small areas, as demonstrated in the top and middle rows.
However, as shown in the bottom row, the result nearly covered the ROI but was slightly misaligned, which led to a false-positive evaluation.

The detection performance of gastric cancer regions from gastric X-ray images is summarized in Table 2.
The object detection model detects regions where the probability (i.e., the confidence score) of belonging to the class of interest exceeds a predefined detection threshold of the bounding box $\xi$ as locations containing objects of that class.
HBBT assigns new negative labels to the over-detected regions that exceed the threshold $\xi$ and retrains the model.
Therefore, to draw an ROC curve for evaluation, the model must be trained every time $\xi$ is changed, which is expensive. 
In this experiment, therefore, the detection threshold $\xi$ was determined such that the SE (recall) of detecting gastric cancer was about 90\%, which was higher than the doctor's 85.5\%.
At this detection threshold, HBBT improved precision by 2.3 points in Faster R-CNN (baseline model 1) and 3.3 points in EfficientDet-D7 (baseline model 2).
Our system, using the proposed techniques, R-sGAIA and HBBT, achieved a recall of 90.2\%, a precision of 42.5\%, and an F1 score of 0.578 (Table 2).

The combination of the proposed R-sGAIA and HBBT showed a significant improvement over RandAugment, one of the most advanced data augmentations, in the detection of gastric cancer using the general object detection model.
Specifically, the improvement was 10.3 points and 6.1 points in precision, and 10.9 points and 5.9 points in F1 scores for the Faster R-CNN (model 1) and the EfficientDet-D7 (model 2), respectively.

The processing time required for one X-ray image was approximately 0.51 seconds, which makes the proposed system a practical and efficient screening tool for gastric X-ray images.

\begin{figure*}[ht]
\begin{center}
\includegraphics[width=130mm]{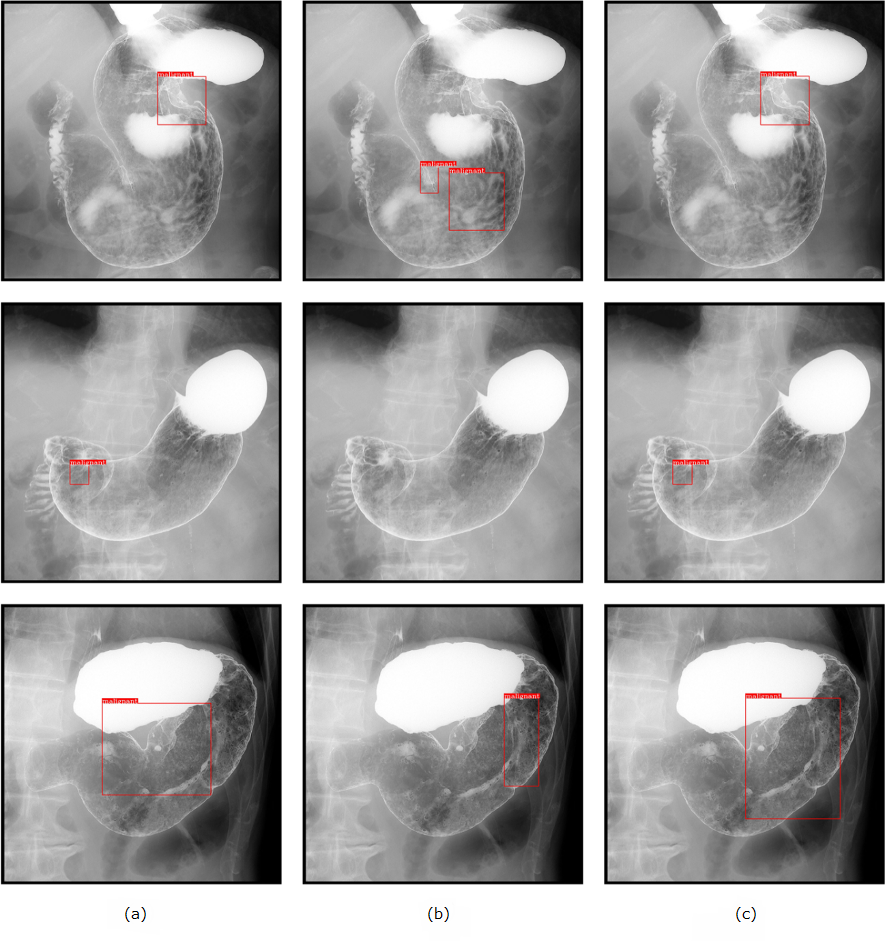}
\caption{Example of cancer detection on EfficientDet-D7 networks (as shown in red box with ``malignant" label): (a) ground-truth region, (b) results obtained with RandAugment (baseline model 2), and (c) results with R-sGAIA + HBBT (proposed). 
The proposed R-sGAIA + HBBT led to the accurate detection for the top and middle examples, whereas for the bottom result, although it showed some improvement, the detection was inaccurate.}
\end{center}
\end{figure*}

\begin{table}[t]
\footnotesize
\begin{center}
\caption{Comparison of the detectability of gastric cancer sites.}
\begin{tabular}{l c c c c c c c} \hline
\multirow{2}{*}{} & \multicolumn{3}{c}{Augmentation} & \multirow{2}{*}{\bf{HBBT}} & Precision & Recall & \multirow{2}{*}{F1 score} \\ \cline{2-4}
  & RandAug ${}^\dag$ & sGAIA ${}^\ddag$ & \bf{R-sGAIA} & & (\%) & (\%) & \\ \hline \hline
\citet{Laddha2019}${}^\ast$ & & & & & 26.5 & 89.7 & 0.409 \\ \hline \hline
Model 1 (Faster R-CNN on ResNet101) & \multicolumn{7}{c}{}  \\ \hline
\quad + RandAugment (baseline model 1) & $\checkmark$ & & & & 28.4 & 90.1 & 0.432 \\ \hline
\quad + sGAIA &  & $\checkmark$ & & & 33.8 & 90.4 & 0.492 \\ \hline
\quad + R-sGAIA &  &  & $\checkmark$ & & 36.4 & 90.2 & 0.519 \\ \hline
\textbf{\quad + R-sGAIA + HBBT (proposed)} &  & & $\checkmark$ & $\checkmark$ & \textbf{38.7} & \textbf{90.1} & \textbf{0.541} \\ \hline \hline
Model 2 (EfficientDet-D7) & \multicolumn{7}{c}{}  \\ \hline
\quad  + RandAugment (baseline model 2)  & 
$\checkmark$ & & & & 36.4 & 90.5 & 0.519 \\ \hline
\quad + sGAIA  & & $\checkmark$ & & & 37.9 & 90.1 & 0.534 \\ \hline
\quad + R-sGAIA & & & $\checkmark$ &  & 39.2 & 90.2 & 0.546 \\ \hline
\quad + HBBT & $\checkmark$ & & & $\checkmark$ & 39.4 & 89.9 & 0.548 \\ \hline
\textbf{\quad + R-sGAIA + HBBT (proposed)} & & & $\checkmark$ & $\checkmark$ & \textbf{42.5} & \textbf{90.2} & \textbf{0.578} \\ \hline
\end{tabular}
\end{center}
${}^\ast$ originally used for gastric polyp detection with YOLOv3 \citep{yolov3}.\\
${}^\dag$ RandAugment \citep{randaugment} with $M = 4$ and $N = 3$. \\
${}^\ddag$ sGAIA \citep{Okamoto2019}. \\

\end{table}

\section{Discussion}
The cancer detection performance of the proposed system in this study was significantly better than that achieved using baseline methods, including RandAugment (a state-of-the-art general data augmentation strategy), existing methods such as sGAIA \citep{Okamoto2019}, and those proposed by \citet{Laddha2019}.
Additionally, the system's cancer detection capability exceeded that of a physician’s reading ($\sim$85.5\%).
In more than two out of five cases, the detection results presented by the system were cancerous.
For a physician using the system, this means that by simply checking the areas surrounded by the few bounding boxes presented in a gastric X-ray image, a higher diagnostic yield can be achieved than with a conventional reading.
In other words, the system is practically effective for on-site medical examinations and can significantly reduce the burden on physicians.

\subsection{R-sGAIA}
The proposed R-sGAIA can probabilistically highlight the gastric folds on a pixel-by-pixel basis based on medical findings, as shown in Figure 5. In this study, we found that R-sGAIA achieved an F1 score improvement of 6.0 points and 2.7 points over RandAugment on baseline models 1 and 2, respectively. This highlights the importance of probabilistic augmentation informed by medical knowledge.
In baseline model 2, R-sGAIA further improved the F1 score by 1.2 points over the previous sGAIA in a direct comparison. This improvement may be attributed to R-sGAIA's use of validation data to determine $p(e)$ as a continuous function, which enables it to generate a wider variety of enhancement images that better highlighted the gastric folds compared to sGAIA, which used a subjectively fixed value.
The function of $p(e)$ with respect to edge strength $e=E(x,y)$ ranged approximately between 0.1 and 0.9, which likely contributed to the performance improvement.
By allowing more stochastic variation in regions with very strong and weak edge strengths, R-sGAIA enhanced the diversity of the generated images, thus leading to better detection performance.

\subsection{HBBT}
Recent state-of-the-art object detection models (e.g., YOLOv10 \citep{yolov10}) are trained to minimize the difference between the boxes specified in the training data and the currently detected boxes, both in terms of their classes (i.e., contents) and locations.
In other words, the model's error is the sum of the class error and position error for each box, and training is performed to minimize these errors.

The proposed HBBT introduces a new and effective strategy that allows learning from healthy control images, which cannot be used to train conventional object detection models. It retrains over-detected boxes by assigning them a unique ``hard sample'' class label to suppress future false detections.
The HBBT strategy increases the class error for error-prone regions (i.e., hard samples) and retrains the model to reduce false positives. In essence, the HBBT learning algorithm is a strategy to reduce generic errors in object detection models, rather than focusing on the specific type of model.
Furthermore, HBBT actively incorporates negative samples, such as images of healthy controls, which are typically excluded from training by conventional object detection models.

Although HBBT is a simple method that recursively learns to eliminate false positives detected in images, including control images, it alone improved precision by 2.3 points and 3.3 points as well as F1 scores by 4.9 points and 2.9 points in baseline models 1 and 2, respectively.
HBBT is a general-purpose learning method that can be combined with any object detection algorithm, not just for gastric cancer detection, as demonstrated in this study.
Because the final model detection is determined by the bounding box confidence threshold, HBBT not only reduces the number of false positives but also helps minimize false negatives (as seen in the middle case in Figure 6).

\subsection{For practical diagnostic support}
In this diagnostic support system, the ability to handle the diversity in size and shape of the tumor area to be detected depends on the detectability of the object detection model used.
The range of object sizes that can be detected is determined by hyperparameters, such as anchor size and aspect ratio.
In the task of cancer detection from gastric X-ray images, the general settings of the object recognition model are sufficient, as the images are taken in a way that ensures the tumor size is large enough for visual diagnosis by a physician.
However, for extremely small tumors, it is necessary to adjust the hyperparameters to account for their size.

Since R-sGAIA is an online data augmentation method and HBBT is a training method, they do not impact the execution time of the diagnostic process.
The current processing time of 0.5 seconds per image is considered sufficiently practical, and further improvements to the detection model are expected to increase speed in the future.

Detecting gastric cancer regions from X-ray images is inherently challenging, not only for automatic diagnostic systems but also for experts.
Despite these challenges, our two technological innovations, R-sGAIA and HBBT, significantly improved the detection performance of gastric cancer, thereby achieving a level of performance and speed that can be effectively used by physicians as a diagnostic aid.
Our diagnostic support system offers a concrete solution to the challenges of gastric radiography, in which it is easy to acquire images but difficult to diagnose.
In the future, we aim to work closely with experts to further refine and improve the system.

\subsection{Comparison with automated diagnostic support methods for other modalities}
Many diagnostic support studies have focused on endoscopy, the most commonly used method for diagnosing gastric cancer.
However, as mentioned earlier, most of these studies did not properly separate the training data from the evaluation data, which led to data leakage and artificially inflated diagnostic performance.
Recent studies that have addressed this issue by using more sophisticated machine learning methods \citep{Shibata2020, Ahmad2023} have reported detection performance for gastric cancer with an image-based F1 score of approximately 0.7, which is considered the current standard for diagnostic accuracy.

Due to its noninvasiveness and ability to image the entire body, X-ray computed tomography (CT) is widely used for detecting not only gastric cancer but also other organs and diseases. Diagnostic studies of gastric cancer using CT typically rely on classical radiomics, including necessary pre-treatment steps. Depending on the cancer stage, diagnostic performance usually achieves an AUC of 0.7 to 0.8 \citep{Meng2020, Huang2022}. The application of deep learning in this context primarily focuses on predicting patient prognosis and segmenting tumor areas rather than directly diagnosing cancer. 
\citet{Hao2022} demonstrated that radiomic features and machine learning-extracted features (i.e., low-dimensional representations) are effective at predicting the prognosis of gastric cancer patients, which indicates that these techniques will become more prevalent in the future.

While the diagnostic performance of these automated support methods may appear lower than the 95\% SE reported by physicians for gastric cancer, it is essential to note that this comparison is patient-based which makes direct comparisons challenging.
As automated diagnostic methods are introduced into clinical practice, their actual performance is likely to approach the current SE of physicians, particularly when multiple outputs are integrated to inform a final decision and provide significant support to physicians.
Although our F1 score of 0.578 for the diagnostic performance of gastric X-ray images may seem lower than that of other modalities, it is important to recognize that this is a box-based evaluation.
Additionally, because physicians diagnose gastric cancer by reviewing multiple images during a gastric X-ray examination, our method’s ability to highlight the ROI in each image makes it highly valuable.
Furthermore, our approach is significant as it represents the first use of gastric X-ray images to aid in the diagnosis of gastric cancer.

\subsection{Limitations of the study}
In recent years, it has become apparent that medical images, such as magnetic resonance (MR) images, contain features unique to each facility due to differences in imaging equipment, protocols, geometry, and other factors.
These so-called domain differences can significantly impact the results of machine-learning tasks, such as classification, when applied without adjustment.
The gastric X-ray images used in this study were only those taken at medical facilities affiliated with Tokai University, and the number of patients (145) and images (4,724) was limited, so the diversity that could be considered was limited.
Differences between institutions and the potential lack of diversity in the training data may affect the analysis of gastric X-ray images taken at different institutions, but this was not verified in this study.
Finally, this experiment did not allow for a detailed discussion of the detection threshold for hard boundary boxes. Addressing this will be an important focus for future work.

\section{Conclusions}
In this paper, we have proposed an unprecedented, accurate, and fast gastric cancer diagnostic support system from gastric X-ray images built on two key innovations: (1) R-sGAIA and (2) HBBT.
The proposed system (R-sGAIA + HBBT on the EfficientDet-D7 network) achieves an SE of 90.2\%, surpassing that of physicians (85.5\%), with a processing time of approximately 0.5 seconds per image.
Notably, 42.5\% of the detected candidate box regions contained cancer, meaning our system effectively narrows down the areas that physicians need to examine.
This system also achieves an F1 score 5.9 points higher than the same network using RandAugment, a state-of-the-art data augmentation method.
Diagnosing gastric cancer using gastric X-rays is a challenging task that requires extensive experience.
We hope that our proposal will assist in diagnosis and lead to early detection and a reduction in the number of deaths from gastric cancer in the near future.

\section*{Acknowledgments}
This research was supported in part by the Ministry of Education, Science, Sports and Culture of Japan (JSPS KAKENHI), Grant-in-Aid for Scientific Research (C), 22K07728, 2022--2024.
No financial support was received that would have influenced the results of this study.

H. Okamoto was responsible for the practical system development of the study and the writing of the paper; 
Q. H. Cap provided model implementation, code publication, and paper writing;
T. Nomura, K. Nabeshima, and J. Hashimoto provided medical guidance, data collection, diagnosis, and other insights; and J. Hashimoto and H. Iyatomi were responsible for the study design, overall management, and writing of the paper.


\bibliographystyle{elsarticle-harv}
\bibliography{main}

\end{document}